\documentclass[11pt, a4paper]{article}

\usepackage{amsthm}
\usepackage{commath}
\usepackage{bbm}
\usepackage{amsfonts}
\usepackage{booktabs}
\usepackage{natbib}
\usepackage{hyperref}
\usepackage{subcaption}
\usepackage{authblk}
\usepackage[export]{adjustbox}
\usepackage{cleveref}
\usepackage[left = 2.5cm, right = 2.5cm, bottom = 3cm, top = 2.5cm]{geometry}

\newcommand*{\bb}{\boldsymbol}

\title{A point process model for rare event detection}

\author[1]{Santhosh Narayanan}
\author[1]{Carsten Maple}
\author[1]{Mark Hooper}

\affil[1]{The Alan Turing Institute}

\begin{document}

\maketitle

\begin{abstract}
Detecting rare events, those defined to give rise to high impact but have a low 
probability of occurring, is a challenge in a number of domains including 
meteorological, environmental, financial and economic. The use of machine 
learning to detect such events is becoming increasingly popular, since they 
offer an effective and scalable solution when compared to traditional 
signature-based detection methods. In this work, we begin by undertaking 
exploratory data analysis, and present techniques that can be used in a 
framework for employing machine learning methods for rare event detection. 
Strategies to deal with the imbalance of classes including the selection of 
performance metrics are also discussed. Despite their popularity, we believe the 
performance of conventional machine learning classifiers could be further 
improved, since they are agnostic to the natural order over time in which the 
events occur. Stochastic processes on the other hand, model sequences of 
events by exploiting their temporal structure such as clustering and dependence 
between the different types of events. We develop a model for classification 
based on Hawkes processes and apply it to a dataset of e-commerce transactions, 
resulting in not only better predictive performance but also deriving inferences 
regarding the temporal dynamics of the data.
\end{abstract}

\section{Introduction}
Many real world applications require the detection of rare observations in large 
data.  For example, intrusions in computer networks \citep{bhuyan2013network}, 
fraudulent transactions in financial data \citep{fiore2019using}, system 
failures \citep{ribeiro2016sequential} and road accidents in traffic data 
\citep{theofilatos2016predicting} are significant but rare events that require 
detection. In the literature, such rare events may be referred to as 
abnormalities, anomalies, novelties, outliers, exceptions, aberrations or 
contaminations, among others.

Rare event detection is recognised as a very challenging problem. It has 
become infeasible for investigators to manually detect rare events in 
real world applications, due to the large number of observations and high 
dimensionality. As a result, the use of machine learning techniques have 
become increasingly popular as they offer an effective and scalable 
solution to automate the process of identifying rare events from 
large volumes of data. However, due to the inherent imbalance of the classes 
in such datasets, statistical methods tend to underestimate the probability 
of rare events and perform poorly \citep{king2001logistic}.
 
In this paper, we present techniques that build towards a general framework 
for employing machine learning based models for rare event detection. Using a 
publicly available dataset of e-commerce transactions, we first illustrate 
how one can develop a better understanding of such data using a series of 
visualisations. Strategies to deal with the imbalance of the classes 
including the crucial selection of performance metrics for evaluating models 
are also discussed.

Typical machine learning algorithms, however, do not naturally account for 
the fact that events are ordered in time and, as such, are unable to capture 
any dependence these events may have on past occurrences. 
Stochastic processes such as Hawkes processes \citep{hawkes1971spectra}, on the 
other hand, model sequences of events by exploiting their temporal structure 
such as clustering and dependence between the different types of events. In 
this paper, we develop a model for classifying events based on Hawkes 
processes that can be applied to a wide range of applications that generate 
event data streams.

We evaluate the accuracy of the point process based classifier by 
comparing against typical ML classifiers and confirm the
superior predictive performance of the point process model. We also 
provide a detailed parameter description showing how the point process model
parameters can be used to gain valuable insights into the underlying 
phenomena that generated the data.

\begin{figure}
\centering
\includegraphics[width = 0.9\textwidth]{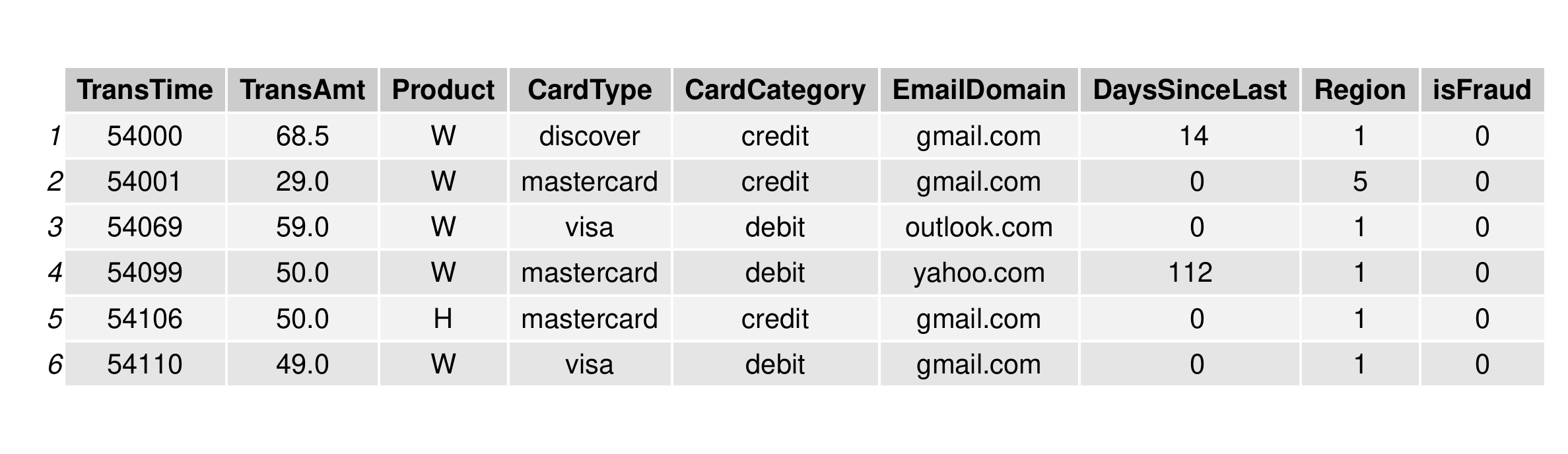}
\caption{Snapshot of the dataset from the Kaggle competition \textit{IEEE-CIS Fraud Detection}.}
\label{fig:snapshot}
\end{figure}

\section{Data}
To illustrate the methodology developed as part of our work, we use a 
public dataset containing real world e-commerce transactions made 
available as part of the Kaggle competition 
\textit{IEEE-CIS Fraud Detection: Can you detect fraud from customer 
transactions?} \citep{kaggledata}. Financial fraud detection is a typical example 
of rare event detection where the data is often characterised by large number 
of samples and a severe imbalance in the number fraudulent samples compared 
to normal ones.

The original dataset contains over 400 variables, however, the vast majority 
of the variables have their names and values masked for privacy protection. 
We exclude the masked variables, as is the case in comparable studies, such as 
\citep{bhattacharyya2011data, carneiro2017data}, since they offer little value 
while illustrating the general modelling framework and its interpretability. 
We retain only nine variables in our reduced dataset including the target variable 
\textsf{isFraud} which is an indicator of the transaction being fraudulent. 
\Cref{fig:snapshot} is a snapshot of the dataset with the variables 
included in our analyses.

\begin{figure}
\centering
\includegraphics[width = 0.4\textwidth]{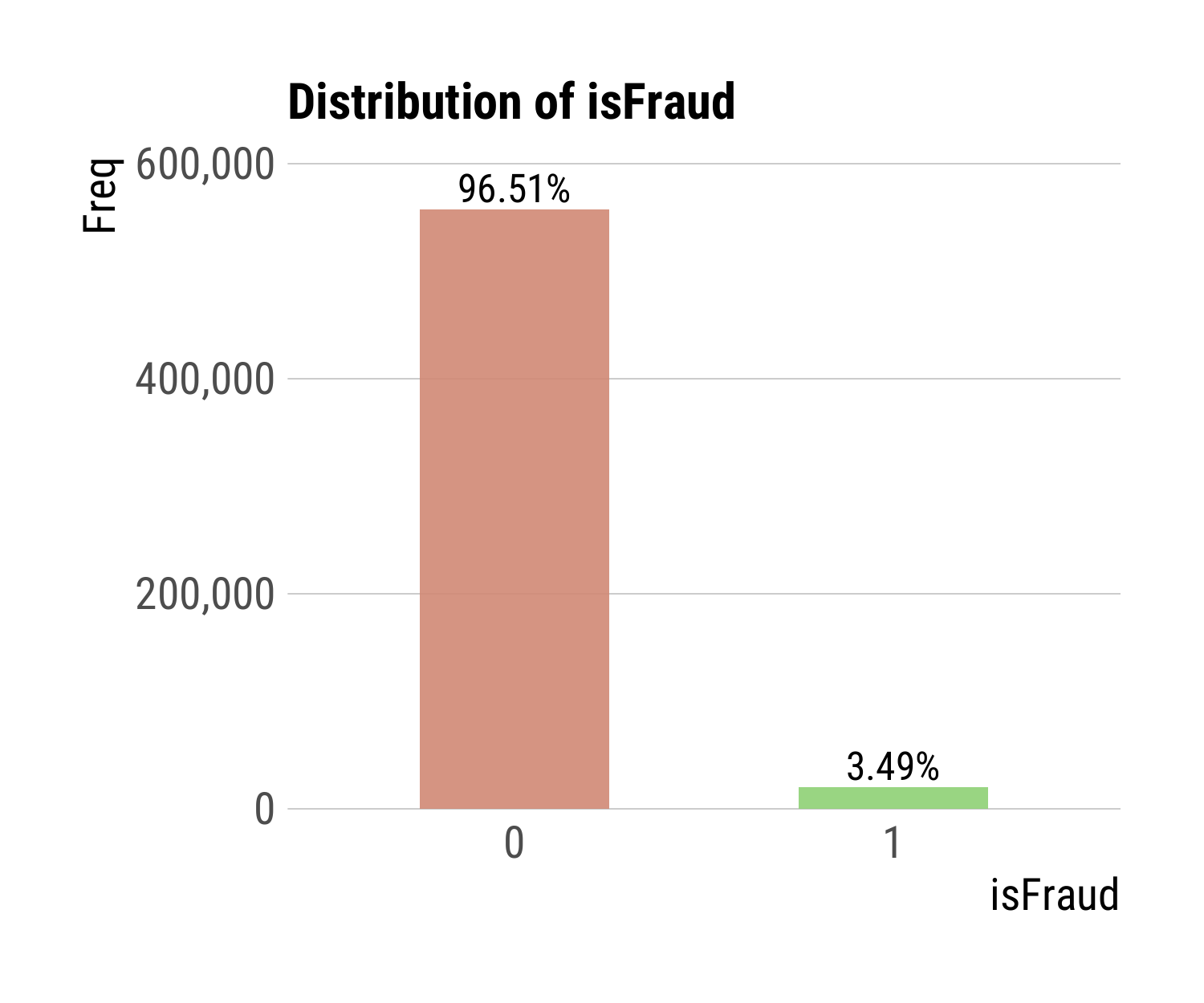}
\caption{Distribution of the target \textsf{isFraud}.}
\label{fig:fraudbarplot}
\end{figure}

\begin{figure}
\centering
\begin{subfigure}{0.45\textwidth}
\includegraphics[width = \textwidth]{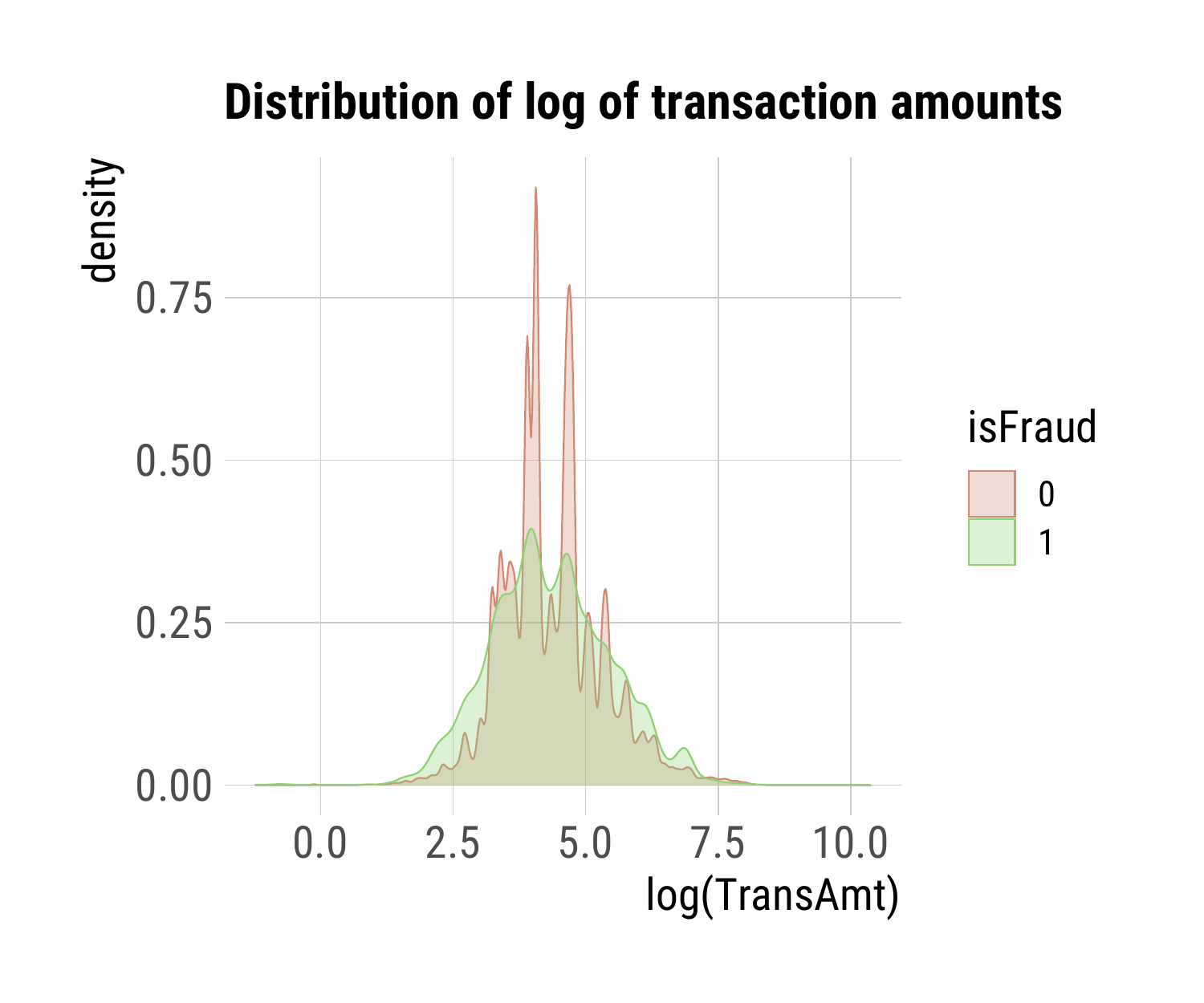}
\caption{Logarithm of the transaction amounts \\ grouped by \textsf{isFraud}.}
\label{fig:amtlogplot}
\end{subfigure}
\begin{subfigure}{0.45\textwidth}
\includegraphics[width = \textwidth]{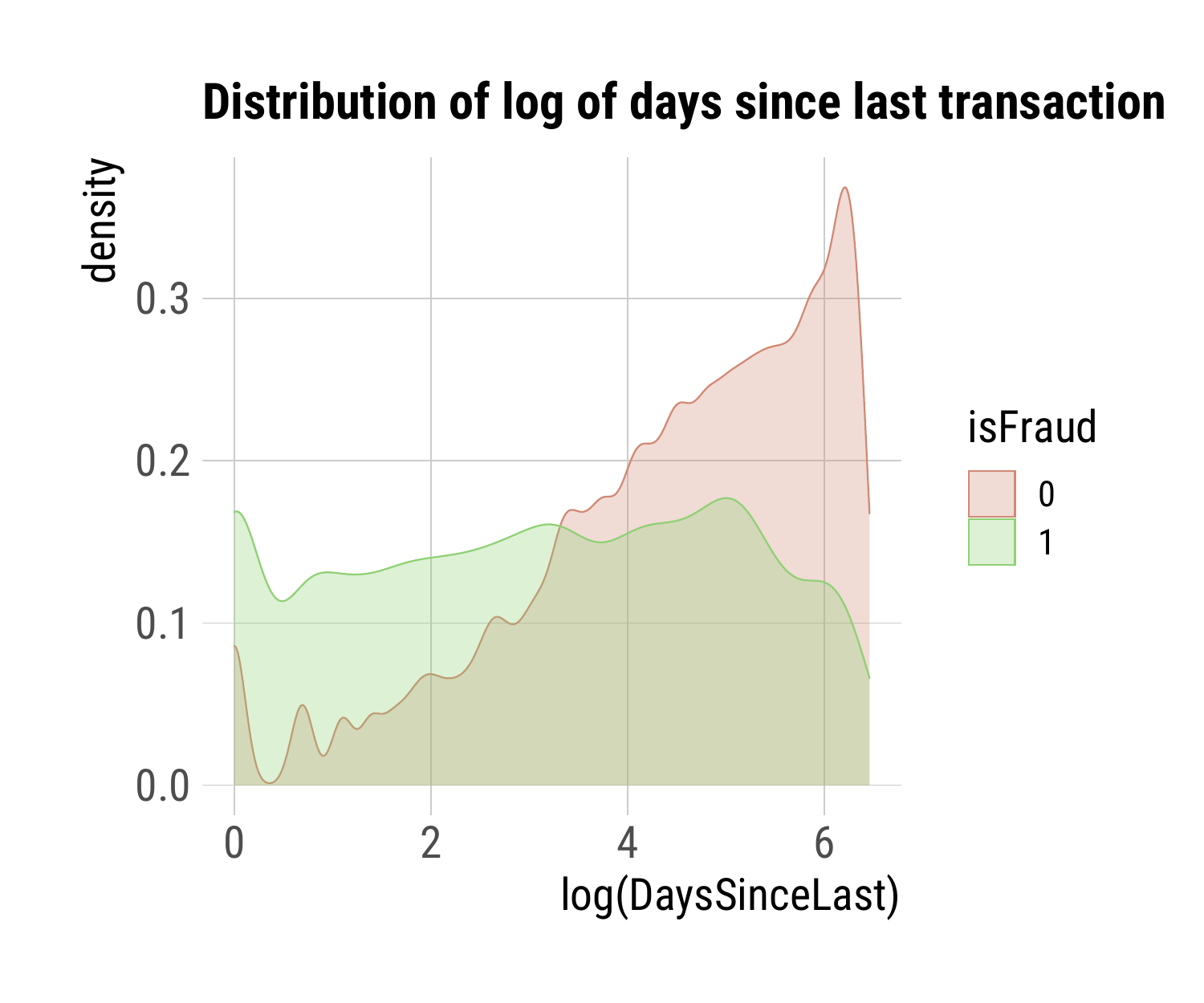}
\caption{Logarithm of the days since last \\ transaction grouped by \textsf{isFraud}.}
\label{fig:dayslogplot}
\end{subfigure}
\caption{Distribution of logarithm of the transaction amounts (left) and the logarithm of the days since the last transaction (right) grouped by \textsf{isFraud}.}
\end{figure}

\begin{figure}
\centering
\begin{subfigure}{0.45\textwidth}
\includegraphics[width = \textwidth]{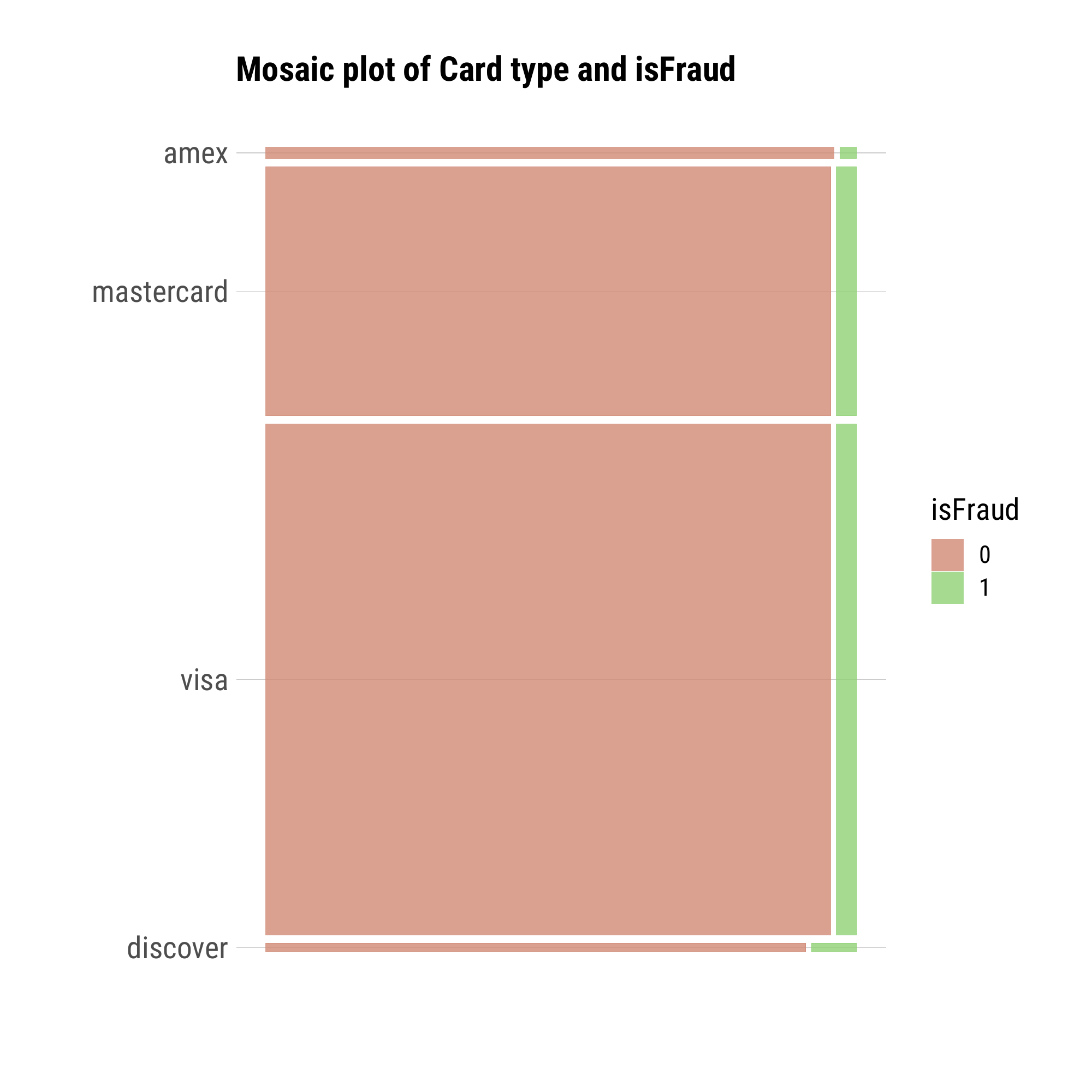}
\caption{\textsf{isFraud} by \textsf{Card type}.}
\label{fig:ctypemosplot}
\end{subfigure}
\begin{subfigure}{0.45\textwidth}
\includegraphics[width = \textwidth]{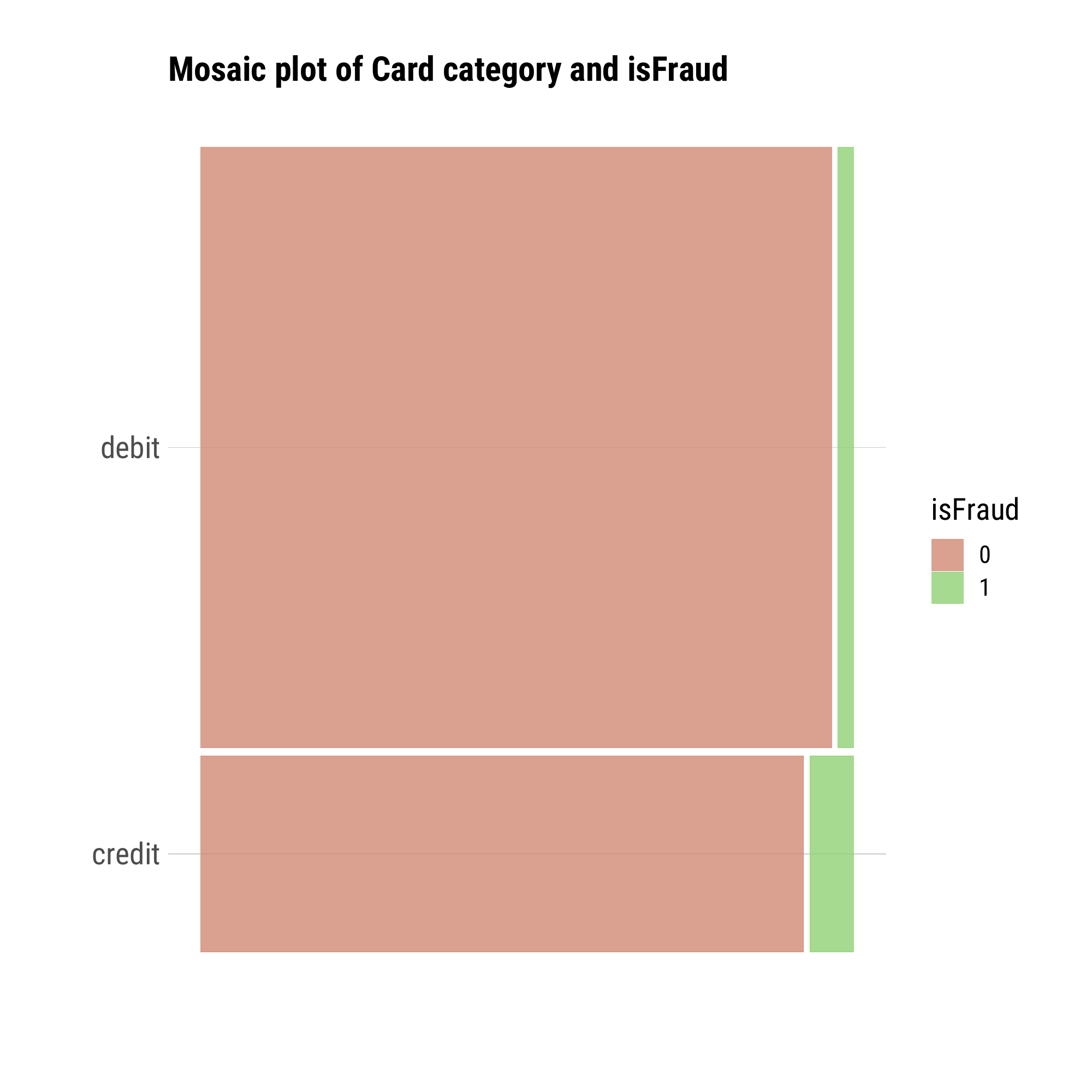}
\caption{\textsf{isFraud} by \textsf{Card category}.}
\label{fig:ccatmosplot}
\end{subfigure}
\caption{Mosaic plots showing the distribution of \textsf{isFraud} by various features.}
\end{figure}

\subsection{Exploratory data analysis}
In this section, we explore the dataset through a series of visualisations to 
get a better understanding of the variables in the data and their 
distributions. \Cref{fig:fraudbarplot} is the distribution of the target 
variable \textsf{isFraud}, showing the severe imbalance in the proportion 
of fraudulent to normal transactions.

\Cref{fig:amtlogplot} is the distribution of the logarithm of the transaction 
amounts grouped by \textsf{isFraud}, where we observe the fraudulent 
transactions having a much smoother distribution over the transaction amounts. 
This could be an indication that fraudsters avoid multiple transactions with 
the same amount.

\Cref{fig:dayslogplot} is the distribution of the logarithm of the days since 
the last transaction grouped by \textsf{isFraud}, where we observe that 
fraudulent transactions are much likely to occur when there has been a 
previous transaction within the last 14 days. This could indicate that 
fraudulent activity is likely to cluster in time.

\Cref{fig:ctypemosplot,fig:ccatmosplot} 
are mosaic plots showing the two-dimensional distribution of the target 
\textsf{isFraud} along with another variable. The variable values along the 
y-axes are ordered with values having the smallest proportion of fraud at 
the top to the largest at the bottom. These figures show the distribution of 
the variable categories as well as help us recognise the categories with 
large fraud proportions.

\section{Related work}
As mentioned earlier, detecting fraudulent financial transactions is a 
typical example of classifying rare events, that is characterised by 
large volumes of data and a severe imbalance of the classes. A systematic 
review of 49 articles using supervised machine learning approaches for fraud 
detection in \citet{ngai2011application} showed that Logistic Regression, 
Decision Trees, Neural Networks and Support Vector Machines are among the 
popular methods of choice. \citet{bhattacharyya2011data} did a comparative 
study of various ML methods for fraud detection under varying proportions of 
fraud during training and report that Random Forests achieved an overall 
higher precision at low recall levels. While most fraud detection techniques 
have been specifically applied to the case of credit-card fraud, 
\citet{carneiro2017data} is one of the few studies that have examined this 
problem in an e-commerce dataset. \citet{carneiro2017data} not only develop 
an automatic system for fraud detection, but also provide insights to fraud 
analysts for improving their manual revision process, which resulted in an 
overall superior performance.

\citet{jensen1997prospective} studied the technical issues in problems like 
fraud detection and identified the severe imbalance of the classes as a key 
challenge. Under-sampling is a strategy to tackle the imbalance problem 
by removing some instances of the majority class in the training set 
\citep{akbani2004applying}, under the assumption that there is redundancy in 
the data. Random over-sampling is another strategy where instances of the 
minority class are replicated \citep{he2009learning}, but its performance is 
limited as it does not add any new information. A more successful approach to 
oversample the minority class is the Synthetic Minority Oversampling Technique 
(SMOTE) \citep{chawla2002smote}, where synthetic examples are generated by 
interpolating between examples of the minority class. 

The selection of an appropriate performance metric is key in imbalanced 
classification problems and it is well-known that measures like Accuracy, 
True Positive Rate and True Negative Rate can be misleading 
\citep{provost2000machine}. As a result, most literature often rely on the area 
under the ROC curve \citep{dal2014learned}, but this ignores the fact that the 
performance on the minority class is more important for businesses.

Almost all methods in the literature for fraud 
detection do not naturally account for the fact that events are ordered in time 
and hence may depend on past occurrences. To the best of our knowledge, the 
development of a classifier based on point processes that model sequences of 
events by exploiting their temporal structure is novel. 

\section{Supervised machine learning approach}
We focus on supervised machine learning methods for rare event 
detection tasks like fraud detection, that can be broken down into two steps. 
The first step of supervised learning consists of building a prediction model 
from a set of labelled (`normal' or `fraudulent') historical data. In the 
second step, the prediction model obtained from the supervised learning process 
is used to predict the label of new transactions.

Formally, a prediction model is a parametric function with parameters 
$\theta$ that takes an input $x$ from an input space 
$\mathcal{X} \in \mathbb{R}^n$, and outputs a prediction 
$\hat{y} = p(x, \theta)$ over an output space $\mathcal{Y} \in \mathbb{R}$.

The input space $\mathcal{X}$ usually differs from the space of raw 
transaction data as most supervised learning algorithms require the input 
domain to be real-valued, and therefore may require the transformation of 
transaction features that are not real numbers (such as timestamps, 
categorical variables, etc). It may also be beneficial to perform feature 
engineering to enrich the transaction data to improve the detection 
performance.

For fraud detection, the output space $\mathcal{Y}$ is usually the predicted 
binary class for a given input $x$, that is $\mathcal{Y} = \{0, 1\}$. 
Alternatively, the output may also be expressed as a fraud probability, with 
$\mathcal{Y} = [0, 1]$, where values closer to 1 express higher probability 
of fraud.

The training of a prediction model $p(x, \theta)$ consists of finding the 
parameters $\theta$ that minimises a loss function, that compares the true 
label $y$ to the predicted label $\hat{y} = p(x, \theta)$ for an input $x$. Due 
to the high class imbalance (much more `normal' than `fraudulent' 
transactions), careful consideration must be given to the choice of loss 
function for fraud detection. Particular care must also be taken in practice 
when splitting the dataset into training and validation sets, due to the 
sequential nature of transactions.

\subsection{Tackling class imbalance}
The challenge of working with imbalanced datasets is that most machine 
learning algorithms will ignore, and in turn have poor performance on, the 
minority class. However, in most cases and particularly for rare event 
detection, it is the performance on the minority class that is most important. 
The popular approaches for tacking class imbalance are based on 
sampling and rely one or both of the following techniques:

\begin{enumerate}
\item \textbf{Over-sampling}, by adding more examples of the minority class 
so it has more effect on the machine learning algorithm. In over-sampling, 
instead of creating exact copies of the minority class examples, we can 
introduce small variations into those copies, creating more diverse synthetic 
samples.
\item \textbf{Under-sampling}, by removing some examples of the majority 
class so it has less effect on the machine learning algorithm. In 
under-sampling we can cluster the examples of the majority class, and do 
the under-sampling by removing examples from each cluster, thus seeking to 
preserve information.
\end{enumerate}

\begin{figure}
\centering
\includegraphics[width = \textwidth]{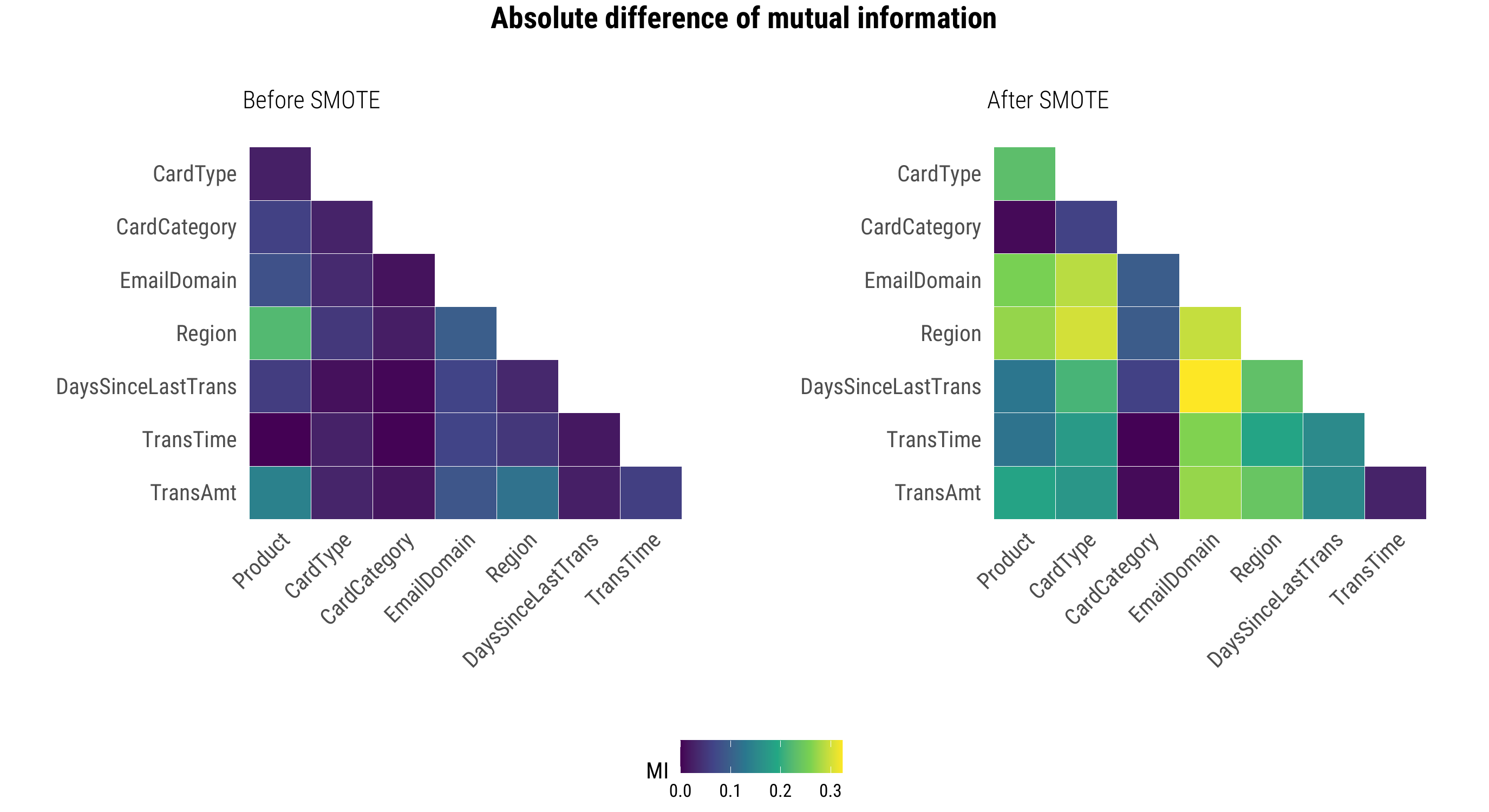}
\caption{The absolute difference of the mutual information (MI) between transactions grouped by \textsf{isFraud} before (left) and after (right) applying the SMOTE technique to balance the classes. The absolute differences of MI are higher after SMOTE indicating that the pairwise feature relationships are more distinct between the two classes after balancing.}
\label{fig:mismoteplot}
\end{figure}

\begin{figure}
\centering
\includegraphics[width = \textwidth]{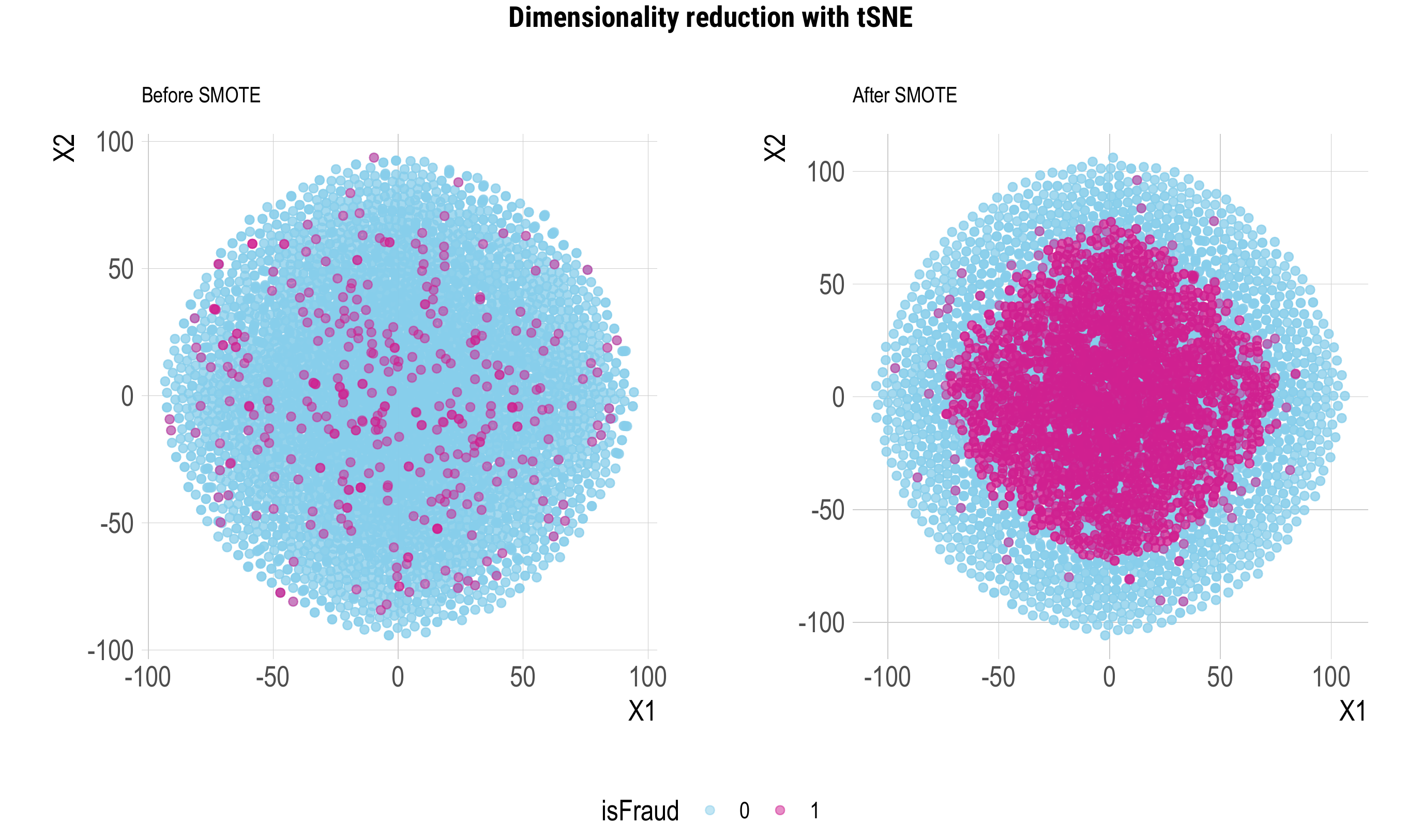}
\caption{Dimensionality reduction using t-SNE before (left) and after (right) using SMOTE to balance the classes. The classes appear to be more separable after the application of SMOTE.}
\label{fig:tsnesmote}
\end{figure}

\subsubsection{SMOTE: Synthetic Minority Over-Sampling Technique}
An effective over-sampling technique where new examples are synthesised 
from the existing examples by introducing small variations is the 
Synthetic Minority Oversampling Technique \citep{chawla2002smote}, 
or SMOTE for short.

SMOTE works by selecting examples that are close in the feature space, 
drawing a line between the examples in the feature space and creating a 
new sample at a point along that line. Specifically, a random example 
from the minority class is first chosen. Then k of the nearest 
neighbours for that example are found (typically k=5). A randomly 
selected neighbour is chosen and a synthetic example is created at a 
randomly selected point between the two examples in feature space.

The approach is effective because new synthetic examples from the 
minority class are created that are plausible as they are relatively 
close in feature space to existing examples from the minority class. 
A general downside of the approach is that synthetic examples are 
created without considering the majority class, possibly resulting 
in ambiguous examples if there is a strong overlap for the classes.

We apply SMOTE to balance the classes in our dataset and assess its 
impact in the following ways. \Cref{fig:mismoteplot} gives the absolute 
difference of the mutual information (MI) among all pairs of features 
between transactions grouped by the target \textsf{isFraud}. The results 
are shown before (top) and after (bottom) the application of SMOTE. The 
absolute differences of MI are comparatively higher after SMOTE 
indicating that the pairwise feature relationships are more distinct 
between the two classes after balancing. 

A popular method for exploring high-dimensional data is t-SNE, introduced 
by \citet{maaten2008visualizing}, an unsupervised and non-linear 
dimensionality reduction technique. \Cref{fig:tsnesmote} gives the 
results of applying t-SNE to visualise the dataset in a reduced 
two-dimensional space both before (top) and after (bottom) using SMOTE to 
balance the classes. The classes appear to be relatively more separable 
after the application of SMOTE.

\begin{table}
\caption{\label{tab:mlrespre}Performance metrics from the modelling experiment before using SMOTE.}
\centering
\begin{tabular}[t]{cccc}
\toprule
Model & Accuracy & ROC\_AUC & PR\_AUC\\
\midrule
Logistic Regression & 0.9582 & 0.7795 & 0.1905\\
Bagged Decision Trees & 0.9524 & 0.6661 & 0.1281\\
\bottomrule
\end{tabular}
\end{table}

\begin{table}
\caption{\label{tab:mlrespost}Performance metrics from the modelling experiment after using SMOTE.}
\centering
\begin{tabular}[t]{cccc}
\toprule
Model & Accuracy & ROC\_AUC & PR\_AUC\\
\midrule
Logistic Regression & 0.8746 & 0.7866 & 0.1898\\
Bagged Decision Trees & 0.9527 & 0.6819 & 0.1358\\
\bottomrule
\end{tabular}
\end{table}

\subsubsection{\label{sec:metric}Choosing the right metric for imbalanced classification}
Selecting an evaluation metric is an important step in any project. 
Choosing the wrong metric can mean choosing a model that solves a different 
problem from the problem we actually want solved.

The metric must capture those characteristics about a model and its 
predictions that are most important to the project stakeholders. This 
is challenging, as there are many metrics to choose from and often project 
stakeholders are not sure what they want. Based on whether the positive 
(minority) class is more important, we make the following recommendations 
for choosing a metric; 

\begin{enumerate}
    \item \textbf{Positive class is more important}
    \begin{itemize}
        \item If both false positives and false negatives are equally important, 
        then use the area under the precision recall curve (PR AUC). This 
        maximises both precision and recall over all possible thresholds.
        \item If false negatives are more costly, then use the F(2) measure. 
        \item If false positives are more costly, then use the F(0.5) measure.
    \end{itemize}
    \item \textbf{Both classes are equally important}
    \begin{itemize}
        \item Use the area under the ROC curve (ROC AUC). This maximises the 
        true positive rate and minimises the false positive rate over all 
        possible thresholds.
    \end{itemize}
\end{enumerate}
where the F($\beta$) measure is defined as
\begin{equation*}
    F(\beta) = \frac{(1 + \beta^2)*\text{precision}*\text{recall}}{\beta^2*\text{precision}*\text{recall}} \, .
\end{equation*}

\Cref{tab:mlrespre,tab:mlrespost} present the performance metrics of the 
machine learning algorithms, logistic regression \citep{kleinbaum2002logistic} 
and bagged decision tree \citep{breiman1996bagging}, for the fraud 
classification task both before and after using SMOTE to balance the classes. 
We use the first 14 days of transactions to train the models and the subsequent 
14 days are used for evaluation. Based on these results it appears that class 
balancing using SMOTE does not provide a significant advantage in classifying 
fraudulent transactions in this dataset.

Despite the rising popularity of machine learning for rare event detection, 
a key limitation suffered by most off-the-shelf machine 
learning methods is that they ignore the natural order in which the events occur 
over time. As it is reasonable to suspect that transactions could be dependent on 
past transactions within the sequence, we explore the idea of building a rare 
event classifier based on point processes.

\section{Point process approach}
The vast majority of machine learning algorithms, with the exception of 
recurrent neural networks, are not designed to model event sequences that 
may have a dependence on past occurrences. In an attempt to overcome this 
limitation, most people resort to clever feature engineering using the 
event timestamp, the success of which is heavily reliant on 
domain expertise as well as the complexity of the underlying phenomena. 
Point processes on the other hand, model sequences 
of events by exploiting their temporal structure like clustering 
and dependence between the different types of events. Such models have the 
potential to not only offer better predictive performance when compared to 
conventional ML classifiers but also provide inferences about the temporal 
dynamics of the data.

Point processes are used to describe a random collection of points in any 
general space, however, we limit ourselves to the case in which the 
points denote events that occur along a time axis. Such temporal 
point processes are well studied \citep[see, for example,][]{daley2003theory} 
and are suitable for a wide range of real-world applications.

Multivariate point processes are those in which two or more types of
points are observed and are specified by associating a random
variable, say $m$ denoting the point type. If $m$ is allowed to
be a general random variable, then we refer to $m$ as a mark and 
the process as a marked point process. An example of a marked point 
process with continuous marks is in seismology, where 
the magnitude of an earthquake is recorded in addition to the time 
of occurrence. In this paper, we model sequences of financial 
transactions using marked point processes with discrete marks 
used to denote whether the transaction is fraudulent or not.

A popular point process model, the Hawkes process, is a mathematical 
model for self-exciting processes proposed in \citet{hawkes1971spectra} 
that can be used to model a sequence of events of some type over time, 
for example, earthquakes. Each event excites the process in the sense that 
the chance of a subsequent event increases for a period of time after 
the initial event and the excitation from previous events add up. 
The marked Hawkes process model captures the magnitudes of all 
cross-excitations between the various event types as well as the 
rate at which these excitations decay over time. Marked Hawkes 
processes are specified using a joint conditional 
intensity function for the occurrence times and the marks 
\citep[see, for example,][expression 2.2]{rasmussen2013bayesian}.

The joint modelling of the times and the marks in a marked point process 
model has to be decoupled, as we are only interested in predicting the 
event type (mark) in our classification model. The joint 
conditional intensity function of a marked point process can be factorised 
into a conditional intensity function of the occurrences times and a 
conditional probability mass function for the event mark. 
\citet{narayanan2021flexible} showed that the conditional 
distribution function for the marks can therefore be derived from the joint 
specification of a marked point process model, resulting in the classification 
model we are interested in.

\begin{figure}
\centering
\begin{subfigure}{0.49\textwidth}
\includegraphics[width=\textwidth]{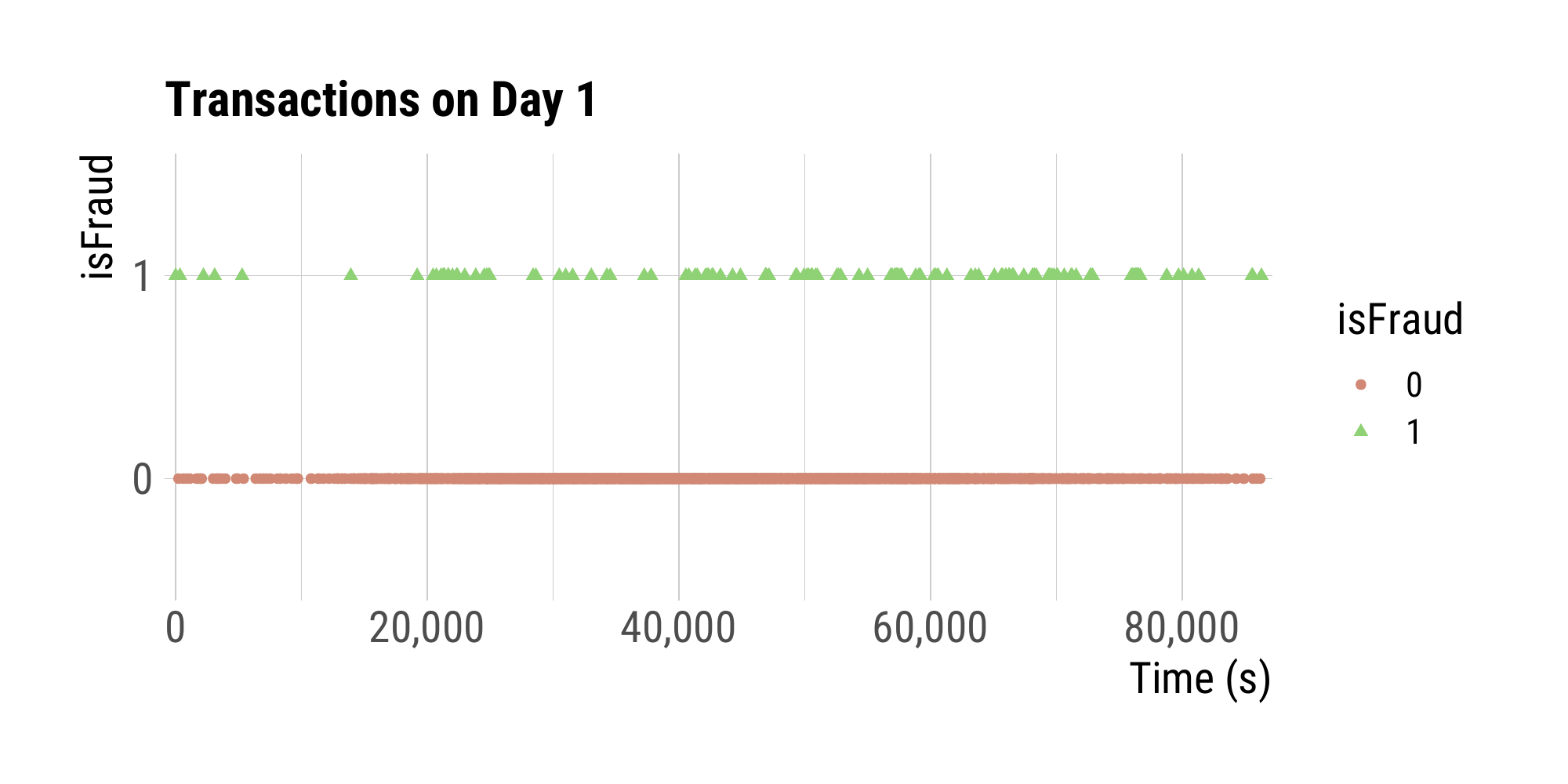}
\caption{Timeline of transactions with isFraud \\indicator.}
\label{fig:basictimeline}
\end{subfigure}
\begin{subfigure}{0.49\textwidth}
\includegraphics[width=\textwidth]{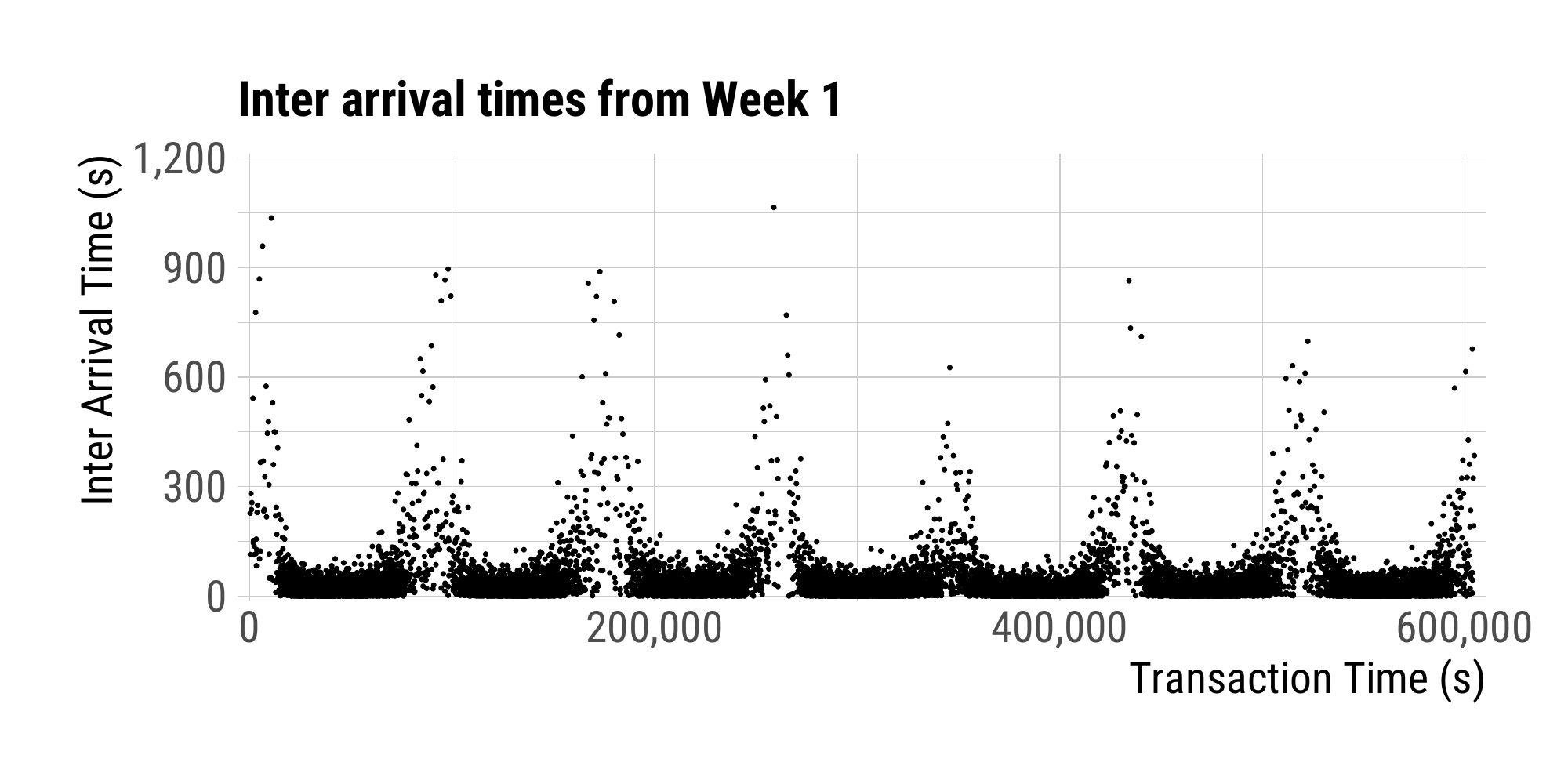}
\caption{Inter-arrival times between transactions \\over the first week.}
\label{fig:intertimes}
\end{subfigure}
\begin{subfigure}{\textwidth}
\centering
\includegraphics[width=0.6\textwidth]{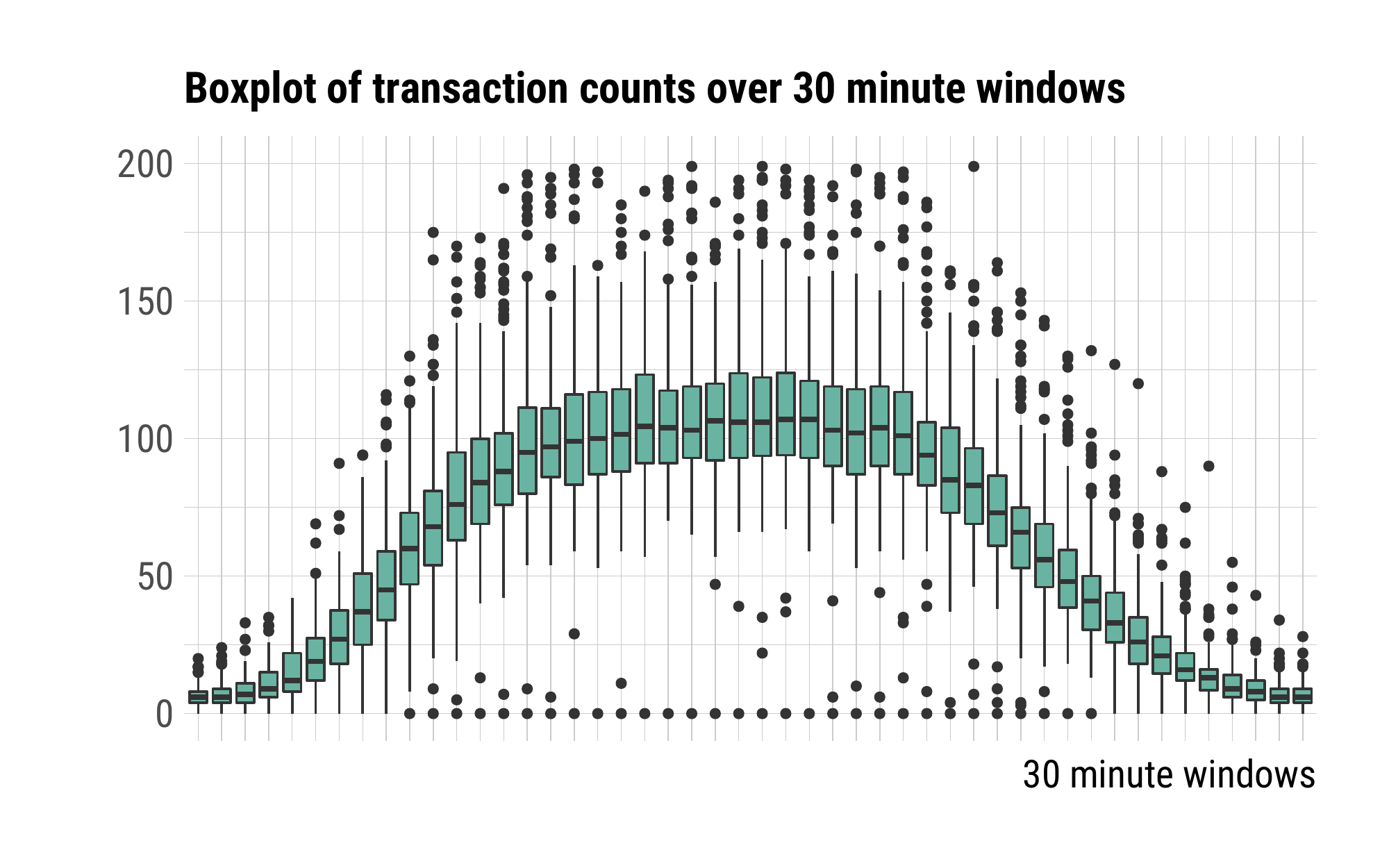}
\caption{Boxplot of transaction counts over the 48 consecutive 30 minute windows over each day.}
\label{fig:boxcounts}
\end{subfigure}
\caption{Exploring the temporal dynamics of the transaction data.}
\end{figure}

\begin{figure}
\centering
\includegraphics[width=0.8\textwidth]{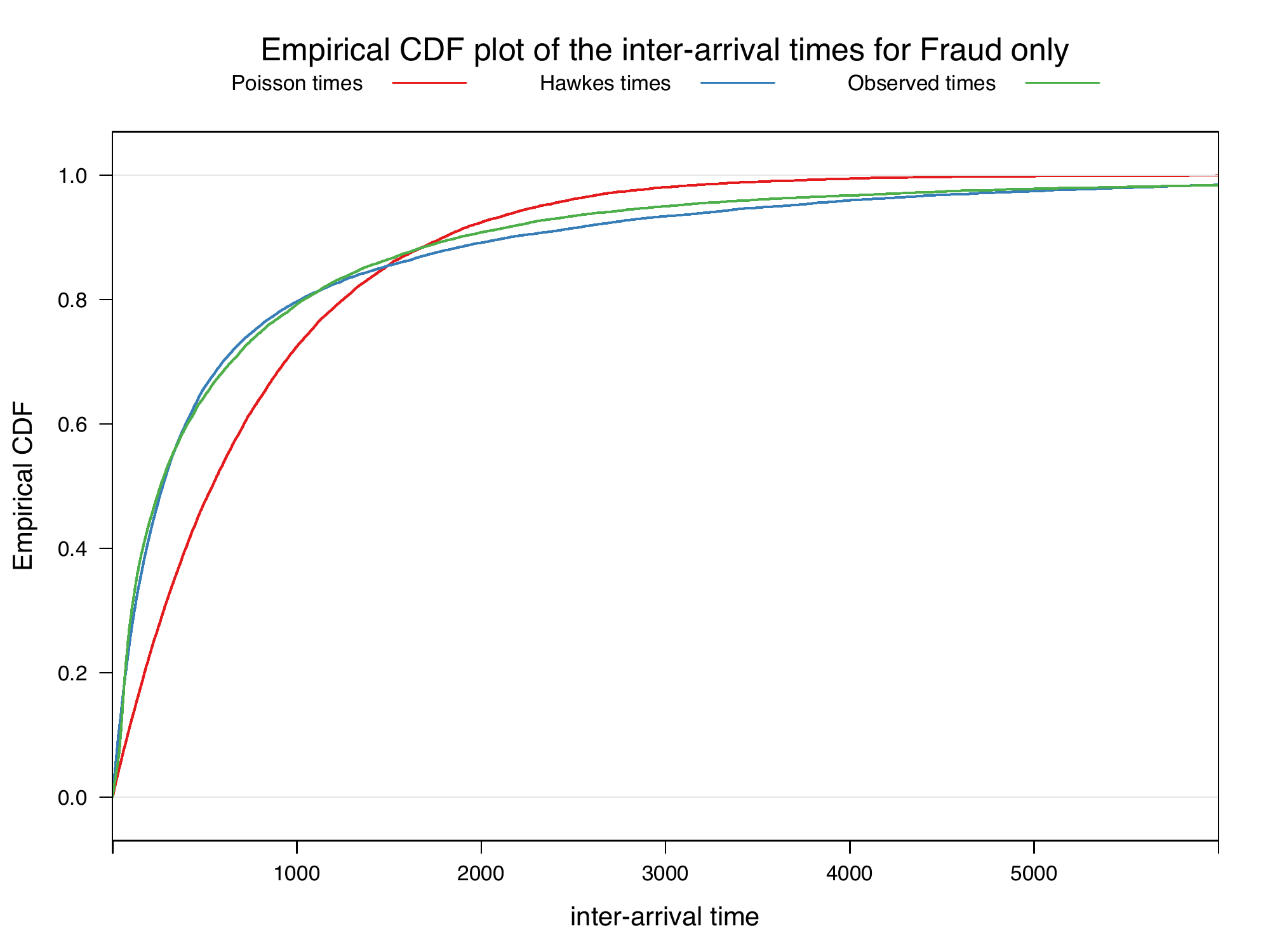}
\caption{Comparing the cumulative distribution functions (CDFs) of the inter-arrival times of events simulated from a Poisson process (red), a Hawkes process (blue), observed event times of the fraudulent transactions.}
\label{fig:ecdfplot}
\end{figure}

\subsection{Exploring temporal dynamics}
The aim here is to investigate the temporal patterns of transactions using 
the theory of point processes. Each transaction is treated as an event 
(point) in the process and the \textsf{isFraud} attribute is the event 
type (mark). The sequence of transactions in the data are therefore 
represented as a marked point process.

\Cref{fig:basictimeline} shows the timeline of transactions with isFraud 
indicator over a single day in the dataset. There is a slight indication 
that fraudulent transactions tend to cluster together. \Cref{fig:intertimes} 
plots the inter-arrival times between transactions over the first week, 
showing a daily seasonality with large inter-arrival times in the early and 
late hours of the day. To further investigate the daily seasonality, in 
\Cref{fig:boxcounts}, we create a boxplot of transaction counts over the 48 
consecutive 30 minute windows that constitute each day in the dataset. 
\Cref{fig:boxcounts} captures the clear pattern in the frequency of 
transactions during the course of a day.

A visual inspection of the occurrence times of fraudulent transactions in
\Cref{fig:basictimeline} offered a slight indication that fraudulent 
transactions may tend to cluster together. To concretely examine if the 
occurrence of a fraudulent transaction increases the probability of 
more fraudulent events to occur soon after, we plot the empirical 
distribution function of the inter-arrival times of the observed 
fraudulent transactions in \Cref{fig:ecdfplot}. We also plot the 
cumulative distribution functions of a homogeneous Poisson process and 
an unmarked Hawkes process, whose parameters are estimated from the 
observed fraudulent events using maximum likelihood. The fitted Poisson 
process, which is memory-less, is far from the empirical distribution 
function, while the fitted Hawkes process provides an excellent fit to the 
observed inter-arrival times. This confirms that the arrival times of fraudulent 
transactions do depend on past occurrences.

\section{Marked point processes}

\subsection{Conditional intensity function}
The collection of the times $\{t_i\}$
at which the events occur and the marks $\{m_i\}$ is a marked point
process, whose ground process, is the process for $\{t_i\}$ only. 
A marked point process is typically specified through its joint conditional 
intensity function
\begin{equation}
\label{exp:mppint}
\lambda^*(t, m) = \lambda^*_g(t)f^*(m \mid t) \, ,
\end{equation}
where $\lambda^*_g(t)$ is the conditional intensity of the ground
process and $f^*(m \mid t)$ is the conditional probability density or
mass function of the mark $m$ at time $t$. Both $\lambda^*_g(t)$ and
$f^*(m \mid t)$ in~\Cref{exp:mppint} are understood as being
conditional on $\mathcal{F}_{t^-}$, which is the filtration of the
marked point process up to but not including $t$.

\subsection{Marked Hawkes processes}

Marked Hawkes processes are point processes whose defining
characteristic is that they {self-excite}, meaning that each arrival
increases the rate of future arrivals for a period of time. More
formally, consider a realisation of a marked point process, consisting
of event times $\{t_i\}$ with $t_i \in \mathbb{R}^+$ and
$t_i > t_{i-1}$, and marks $m_i \in \{1, \ldots, M\}$
$(i = 1, \dots, n)$, where $M$ is the number of discrete marks. The
marked Hawkes process is most intuitively specified using its mark
dependent conditional intensity function $\lambda^*(t, m)$, which for
an exponentially decaying intensity is \citep[expression
2.2]{rasmussen2013bayesian},
\begin{equation}
\lambda^*(t, m) = \mu\delta_{m} + \sum_{t_j < t} \alpha\beta \mathrm{e}^{-\beta (t-t_j)}\gamma_{m_{j} \to m} \, .
\label{exp:hawkint}
\end{equation}
In~\Cref{exp:hawkint}, the parameter $\mu > 0$ is a constant
background intensity and $\delta_{m} \in (0,1)$ is the background mark
probability for mark $m$ with $\sum_{m = 1}^M \delta_m = 1$. The
parameter $\alpha \in (0,1)$ is the excitation factor, $\beta > 0$ is
the exponential decay rate and $\gamma_{m_{j} \to m} \in (0,1)$ is the
probability the excitation from an event of mark $m_j$ triggers an
event of mark $m$, with $\sum_{m = 1}^M \gamma_{m_{j} \to m} = 1$ for
any $m_j \in \{1, \ldots, M\}$.

\subsection{Deriving the classification model}
The conditional probability distribution function of the marks 
$f(m_i \mid t_i, \mathcal{F}_{t_{i-1}}; \bb{\theta})$, where $\bb{\theta}$ 
is an unknown parameter vector, is the classification model we wish to use.
As shown in \citet{narayanan2021flexible}, 
$f(m_i \mid t_i, \mathcal{F}_{t_{i-1}}; \bb{\theta})$ can be derived from 
the joint conditional intensity function of the marked Hawkes process model 
in~\Cref{exp:hawkint} as follows.

By the definition of the conditional intensity function for a marked
point process in~\Cref{exp:mppint},
$f(m_i \mid t_i, \mathcal{F}_{t_{i-1}}; \bb{\theta}) = \lambda^*(t_i,
m_i) / \sum_{m = 1}^M \lambda^*(t_i, m)$. Plugging in $\lambda^*(t_i, m)$
from~\Cref{exp:hawkint} in the latter expression, gives
\begin{equation}
f(m_i \mid t_i, \mathcal{F}_{t_{i-1}}; \bb{\theta}) = \frac{\delta_{m_i} + \sum_{t_j < t_i} \alpha^* \mathrm{e}^{- \beta (t_i-t_j)}\gamma_{m_{j} \to m_{i}}}{1 + \sum_{t_j < t_i} \alpha^* \mathrm{e}^{- \beta (t_i-t_j)}} \, ,
\label{exp:modspec}
\end{equation}
where $\alpha^* = \frac{\alpha\beta}{\mu}$. The 
parameters  $\mu$ and $\alpha$ of the marked Hawkes process as specified 
by~\Cref{exp:hawkint} are not identifiable. Apart from a mathematical 
fact, this is also rather intuitive, because $\mu$ and 
$\alpha$ in~\Cref{exp:hawkint} characterise the evolution of the Hawkes 
process in the time dimension and the sequence of marks is not sufficient 
to identify them.

\subsection{Parameter interpretation}
\label{sec:modparinterpret}

In~\Cref{exp:modspec}, the mark probability of each event in the 
sequence is determined by a combined additive effect from a background 
component and all previous occurrences. The first term 
$\delta_{m_i} \in (0,1)$ in the numerator is the probability 
an event has a mark $m_i$ if the event is triggered solely by the background 
component. Each term 
$\alpha^* \mathrm{e}^{- \beta (t_i-t_j)}\gamma_{m_{j} \to m_{i}}$ is the 
contribution from a previous occurrence in the 
sequence while the excitation factor $\alpha^* \geq 0$ is a scaling factor 
applied to the contributions from the previous occurrences. Large values of 
$\alpha^*$ indicate a stronger 
dependence of the process on its history, since the contributions from 
previous occurrences are weighted higher relative to the background 
component. The decay rate $\beta > 0$ is the exponential rate at which 
the excitations from previous occurrences decay over time. The parameter 
$\gamma_{m_{j} \to m_i} \in (0, 1)$ is the probability the excitation from 
an event of mark $m_j$ triggers an event of mark $m_i$. In other words, 
$\gamma_{m_{j} \to m_i}$ can be viewed as the conversion rate for the 
transition from an event with mark $m_{j}$ to an event with mark $m_i$.

In summary, as in marked Hawkes processes, the specification for the marks 
in~\Cref{exp:modspec} captures not only all cross-excitations between 
the various marks but also the rate at which these excitations decay over 
time.

\subsection{Including covariates}
Transaction datasets in finance and other related sectors, typically carry 
additional covariates apart from the occurrence times and the indicator 
variable for fraud. In such cases, we can allow the background mark 
probability $\bb{\delta}$ and event conversion rates $\bb{\gamma}$ to depend 
on the covariate as
\begin{equation}
\label{exp:modspeccov}
f(m_i \mid t_i, z_i, \mathcal{F}_{t_{i-1}}; \bb{\theta}) = \frac{\delta_{m_i \mid z_i} + \sum_{t_j < t_i} \alpha^* \mathrm{e}^{ - \beta (t_i-t_j)}\gamma_{m_{j} \to m_{i} \mid z_i}}{ \sum_{m=1}^M \left[ {\delta_{m \mid z_i} + \sum_{t_j < t_i} \alpha^* \mathrm{e}^{- \beta (t_i-t_j)}\gamma_{m_{j} \to m \mid z_i}} \right]} \, .
\end{equation}
where $\{z_i\}$ is the collection of the covariate component of the process 
and the filtration $\mathcal{F}_{t_{i}}$ now includes all times, marks and 
covariates up to time $t_{i}$.

\begin{table}
\caption{\label{tab:counts}Aggregated counts of event types over each of the 14 days in the training dataset. `Normal' transactions are coded as (mark $=$ 1) and `fraudulent' ones coded as (mark $=$ 2).}
\centering
\begin{tabular}[t]{ccc}
\toprule
day & mark = 1 & mark = 2\\
\midrule
1 & 2462 & 111\\
2 & 2235 & 86\\
3 & 2684 & 127\\
4 & 2819 & 99\\
5 & 2900 & 117\\
6 & 2838 & 141\\
7 & 3131 & 148\\
\bottomrule
\end{tabular}
\begin{tabular}[t]{ccc}
\toprule
day & mark = 1 & mark = 2\\
\midrule
8 & 2409 & 152\\
9 & 2283 & 95\\
10 & 2846 & 78\\
11 & 2581 & 116\\
12 & 2671 & 105\\
13 & 2557 & 118\\
14 & 3156 & 142\\
\bottomrule
\end{tabular}
\end{table}

\section{Results}
We use the specification in \Cref{exp:modspeccov} as the model that 
classifies the transactions in our dataset as `normal' or `fraudulent'. We use 
only three out of the nine variables from the dataset as shown in 
\Cref{fig:snapshot}, namely the \textsf{TransTime} variable as the time 
dimension, the \textsf{Product} variable as the covariate and the 
\textsf{isFraud} indicator as the mark dimension.

\Cref{tab:counts} shows the aggregated counts of event types over each of the 14 
days that constitute the training dataset. `Normal' transactions are coded as 
(mark $=$ 1) and `fraudulent' ones coded as (mark $=$ 2). The five types of 
products are also encoded as integers from 1 to 5.

\begin{table}
\caption{\label{tab:mle}Maximum likelihood estimates with 95\% CI.}
\centering
\begin{tabular}[t]{cccc}
\toprule
parameter & value & lower & upper\\
\midrule
\addlinespace
$\delta_{1 \mid 1}$ & 0.5498 & 0.3161 & 0.7635\\
\addlinespace
$\delta_{1 \mid 2}$ & 0.2790 & 0.0326 & 0.8161\\
\addlinespace
$\delta_{1 \mid 3}$ & 0.6599 & 0.2521 & 0.9179\\
\addlinespace
$\delta_{1 \mid 4}$ & 1.0000 & 1.0000 & 1.0000\\
\addlinespace
$\delta_{1 \mid 5}$ & 0.9438 & 0.8954 & 0.9705\\ [0.1cm]
\addlinespace
$\alpha$ & 3.7011 & 1.6297 & 8.4051\\
\addlinespace
$\beta$ & 0.2649 & 0.2146 & 0.3269\\ [0.125cm]
\bottomrule
\end{tabular}
\begin{tabular}[t]{cccc}
\toprule
parameter & value & lower & upper\\
\midrule
$\gamma_{1 \to 1 \mid 1}$ & 0.9308 & 0.9159 & 0.9432\\
$\gamma_{2 \to 1 \mid 1}$ & 0.0000 & 0.0000 & 0.0000\\
$\gamma_{1 \to 1 \mid 2}$ & 0.9466 & 0.9204 & 0.9645\\
$\gamma_{2 \to 1 \mid 2}$ & 0.2661 & 0.0714 & 0.6309\\
$\gamma_{1 \to 1 \mid 3}$ & 0.9538 & 0.9337 & 0.9680\\
$\gamma_{2 \to 1 \mid 3}$ & 0.0000 & 0.0000 & 0.0000\\
$\gamma_{1 \to 1 \mid 4}$ & 0.9422 & 0.9160 & 0.9606\\
$\gamma_{2 \to 1 \mid 4}$ & 0.0000 & 0.0000 & 0.0000\\
$\gamma_{1 \to 1 \mid 5}$ & 0.9860 & 0.9840 & 0.9878\\
$\gamma_{2 \to 1 \mid 5}$ & 0.9348 & 0.8945 & 0.9604\\
\bottomrule
\end{tabular}
\end{table}

\Cref{tab:mle} gives the 
parameter estimates including their 95\% confidence intervals after performing 
parameter estimation using maximum likelihood.
The excitation factor $\alpha$ in \Cref{exp:modspeccov} 
is a scaling factor applied to the contributions from the previous 
occurrences to the event mark probability. In \Cref{exp:modspeccov}, 
the background component has a weight of 1, while previous 
occurrences are weighted by $\alpha$. The 95\% confidence interval for 
$\alpha$ is $(1.63, 8.41)$, meaning the contributions from 
previous occurrences carry higher weight relative to the background 
component. In other words, this indicates that financial transactions
have a significant dependence on their history.

The decay rate $\beta$ is the exponential rate at which the excitations 
from previous occurrences decay over time. The 95\% confidence interval for 
$\beta$ is $(0.21, 0.33)$, meaning the excitation effect caused by an event
lasts for at least 44 minutes after its occurrence before decaying 
(contribution drops less than $10^{-6}$).

The background mark probability $\delta_{m \mid z}$ in \Cref{tab:mle} 
is the probability an event with covariate $z$ has a mark $m$ if the event
is triggered solely by the background component. An interesting case is of 
that of product code 4 where the background probability for a fraudulent 
transaction is nearly 0. In other words, if a fraudulent transaction occurs with 
product code 4 it has to be necessarily triggered by one of the past 
occurrences.

\begin{table}
\caption{\label{tab:gammamle} Maximum likelihood estimates of the triggering probabilities.}
\centering
\begin{tabular}[t]{crcc}
\toprule
product &  $\gamma_{r \to c}$ & 1 & 2 \\
\midrule
1 & 1 & 0.9308 & 0.0692\\
1 & 2 & 0.0000 & 1.0000\\
\addlinespace
2 & 1 & 0.9466 & 0.0534\\
2 & 2 & 0.2661 & 0.7339\\
\addlinespace
3 & 1 & 0.9538 & 0.0462\\
3 & 2 & 0.0000 & 1.0000\\
\addlinespace
4 & 1 & 0.9422 & 0.0578\\
4 & 2 & 0.0000 & 1.0000\\
\addlinespace
5 & 1 & 0.9860 & 0.0140\\
5 & 2 & 0.9348 & 0.0652\\
\bottomrule
\end{tabular}
\end{table}

The triggering probability $\gamma_{r \to c}$ in \Cref{tab:gammamle} 
is the probability the excitation from an event of mark $r$ triggers an 
event of mark $c$. The diagonal elements of the $\gamma$ matrix are close 
to 1 for all products with the exception of product code 5, indicating that 
events almost exclusively trigger more events of the same type. In other words, 
for example, the occurrence of a fraudulent event is highly likely to trigger 
another fraudulent event as compared to a normal event.

In summary, the parameters of the model provide valuable insight into 
the underlying processes that generated the data. Specifically, we learned 
how strongly financial transactions depend on their history, the range or 
duration over which the effects of past events persist as well as the 
magnitude of cross-excitations that capture the dependence between the 
different types of events.

\begin{figure}
\centering
\includegraphics[width = 0.7\textwidth]{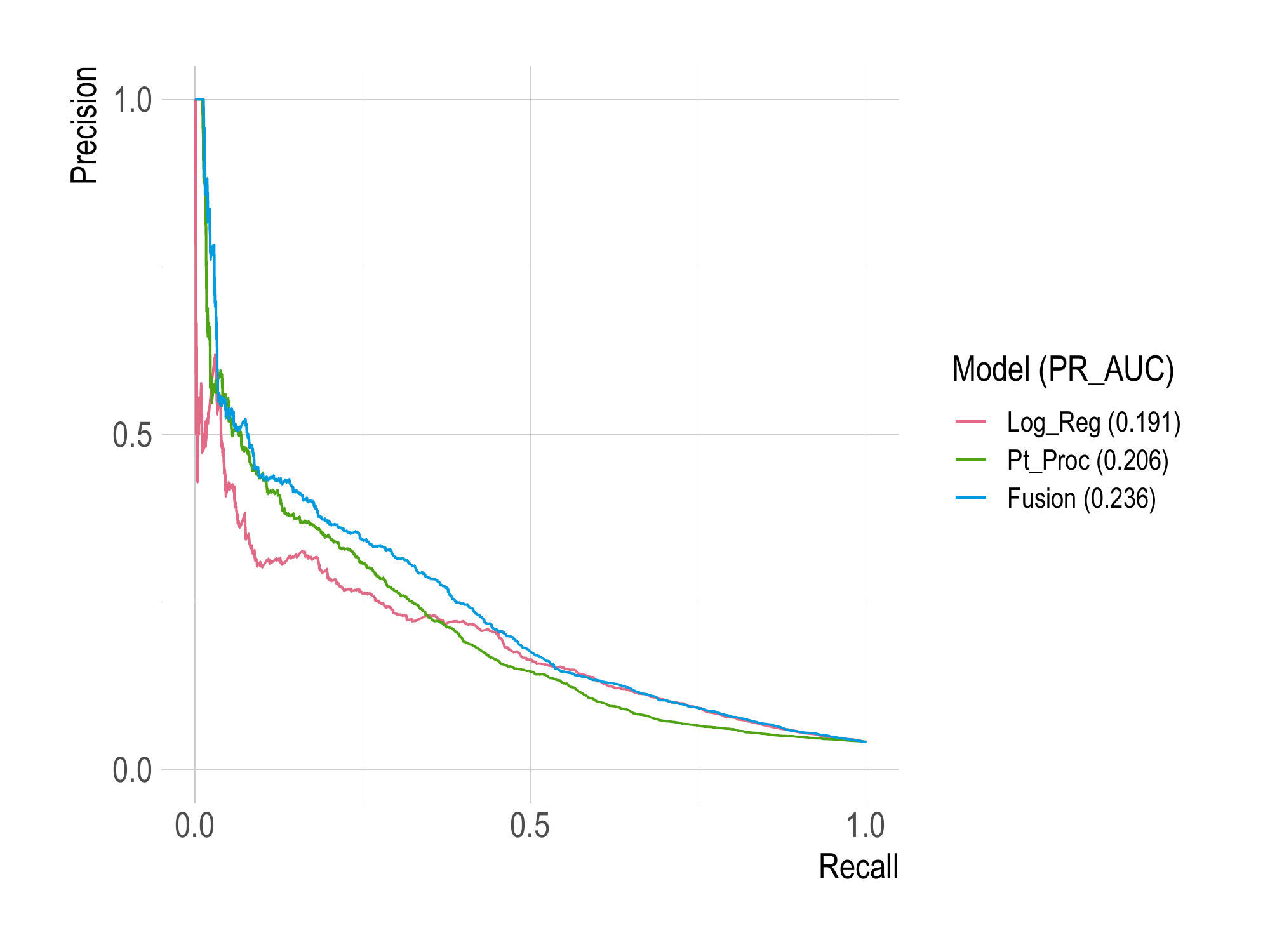}
\caption{\label{fig:prc}Area under the Presicion-Recall Curve.}
\end{figure}

\Cref{fig:prc} evaluates the predictive performance of the point 
process based model on the test set which contains transactions 
over the 14 days immediately after the training data. We use the 
area under the Presicion-Recall curve as the metric for evaluation, 
which treats the positive class (fraudulent) as more important and 
weights both false positives and false negatives equally as discussed 
in \cref{sec:metric}. We also include the logistic regression model 
from \Cref{tab:mlrespre} for comparison, which is a typical example 
of an ML algorithm that does not capture temporal dependence, and we 
see that the point process model indeed performs better.

\section{Conclusions}
The performance of the point process model is remarkable despite using
just two features from the dataset, namely, the occurrence times and the 
product covariate, to classify transactions as normal or fraudulent. In 
contrast, the logistic regression model uses all of the eight features 
available in the data shown in \Cref{fig:snapshot}. We also present a fusion 
model in \Cref{fig:prc}, derived by averaging the predicted probabilities 
from the two fitted models, in an attempt to combine their arguably 
different strengths. The fusion model does perform better than either of the 
models and the fact that even such a naive fusion works well suggests that 
the two models presented indeed capture very different patterns in the data.

\begin{figure}
\centering
\includegraphics[width = 0.9\textwidth]{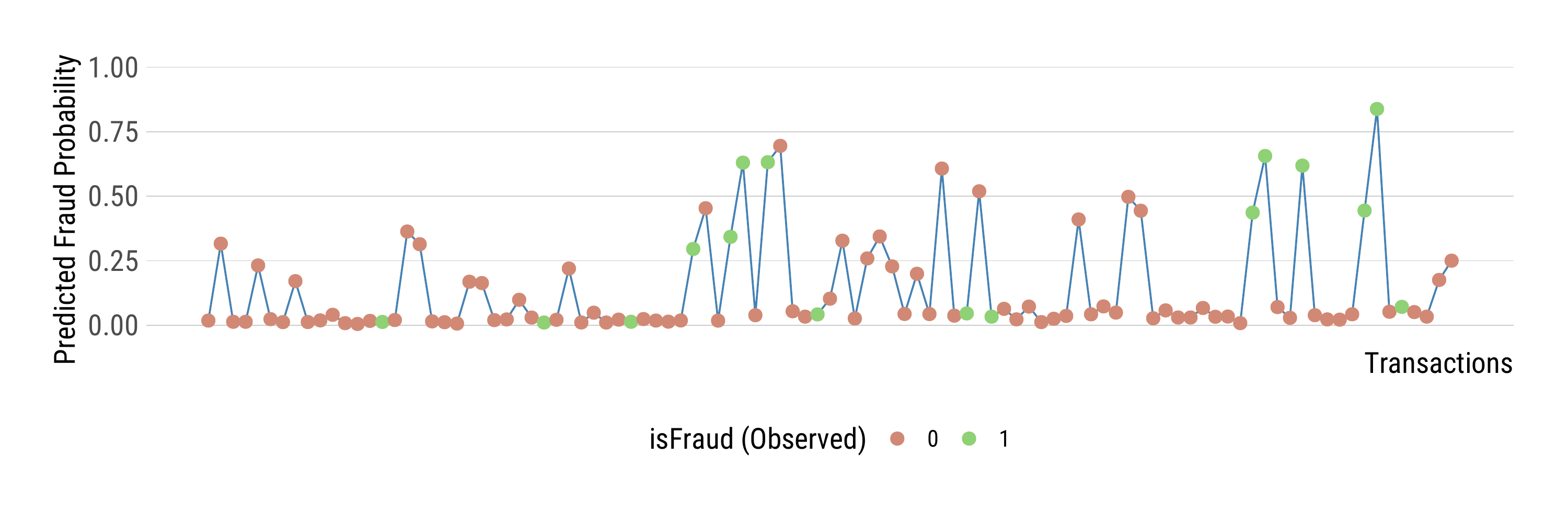}
\caption{\label{fig:pfp}Timeline of predicted fraud probabilities of the first 100 transactions in the test set.}
\end{figure}

In \Cref{fig:pfp} we show a timeline of predicted fraud probabilities 
from the point process model, for the first 100 transactions in the test set, to 
demonstrate how such models can be used in practise. As expected, the 
model with excitation effects is able to capture the noticeable clustering of 
fraudulent transactions.

None of the methodology used in our analyses have been tailored specifically 
for fraud detection or even rare event detection for that matter. Hence, they 
can be applied to a wide range of applications that generate event data streams.

\bibliographystyle{chicago}
\setcitestyle{authoryear,open={(},close={)}}
\bibliography{ref}

\end{document}